\documentstyle[prl,preprint,aps,epsf]{revtex}

\tightenlines

\begin{document}

\draft
\title{ A Paradox in the Langevin Equation with Long-Time Noise Correlations}

\author
{T. Srokowski}

\address{
 Institute of Nuclear Physics, PL -- 31-342 Krak\'ow, Poland }

\date{\today}

\maketitle

\begin{abstract}
\parbox{14cm}{\rm 
We solve the generalized Langevin equation driven by a stochastic force
with power-law autocorrelation function. A stationary Markov process has been 
applied as a model of the noise. However, the resulting velocity variance 
does not stabilizes but diminishes with time. It is shown that algebraic 
distributions can induce such non-stationary affects. Results are compared 
to those obtained with a deterministic random force. Consequences for 
the diffusion process are also discussed. }

\end{abstract}

\vskip2cm

\vfill
\newpage

\narrowtext

Modeling a physical system in terms of Langevin formalism must take into
account the nature and origin of the stochastic force. Usually that force is 
taken in the form of the white noise but in many cases that is an 
unrealistic idealization. Among the systems possessing a finite noise
correlation time, those with power-law (algebraic) correlations are especially
interesting because of lack of characteristic time scale and divergent moments.
Such systems are not unusual. The algebraic random force autocorrelation 
function (FAF) appears in the fluid dynamics \cite{maz,cho} and linearized
hydrodynamics \cite{cic}. For such phenomena as the noise-induced Stark 
broadening \cite{frisch1} and nuclear collisions \cite{sron}, correlation 
functions proportional to $1/t$ have been found. The latest form of 
correlations is of special importance for molecular dynamics because it 
corresponds to the problem of scattering inside a periodic lattice \cite{zach}.

For systems with finite noise correlation time, the ordinary Langevin equation 
must be generalized \cite{kubo} to ensure proper fluctuation-dissipation 
relations. In the absence of external potential, this equation has the form
\begin{eqnarray}
\label{gle}
m \frac{dv(t)}{dt} = -m\int_0^t K(t-\tau)v(\tau)d\tau + F(t)~~~~\ 
\end{eqnarray}
where $F(t)$ is a stochastic force and $K(t)$ denotes the retarded friction 
kernel. The fluctuation-dissipation theorem imposes the relation \cite{kubo1}: 
$K(t)=\langle F(0)F(t)\rangle /mT, $
with temperature $T$ and mass $m$. We require $\langle F(t)\rangle=0$ 
and the following FAF:
\begin{eqnarray}
\label{cor}
C_F(t)\equiv \langle F(0)F(t)\rangle=\cases 
{\beta T/\epsilon~~~&for$~~~~t\le \epsilon$\cr
\beta T/t~~~&for$~~~~t>\epsilon$}
\end{eqnarray}
where $\epsilon$ is a small number. The coefficient $\beta$ we set equal to
one. The solution of Eq.(\ref{gle}) with the initial condition $v(0)=0$
takes the form \cite{glang}
\begin{equation}
\label{solv}
v(t)=g(t)+\int_0^t R(t-\tau)\,g(\tau)\,d\tau
\end{equation}
where $g(t)=m^{-1}\int_0^t F(\tau)\,d\tau$ and the resolvent is given by
$R(t)=\exp(-at)\,(c_1\sin bt +c_2\cos bt)+
m \int_0^\infty x\exp(-tx)/[(mx+\mbox{Ei}_1(\epsilon x)-1)^2+\pi^2]\;dx$. 
The modified
integral exponential function Ei$_1(x)$ is defined by the series:
Ei$_1(x)=\gamma +\ln x +\sum_{n=1}^\infty x^n/n!\,n,$
where $\gamma=0.5772157\dots\;$ is the Euler constant. The other constants 
are fixed in the following at the values: 
$a=-3.52832$, $c_1=-4.76673$, $c_2=-5.35498$ and $b=2.49975$, corresponding 
to $\epsilon=0.01$ and $m=1$.

We are interested in evaluation of the second moment of velocity distribution
$\langle v^2\rangle(t)$. It can be obtained directly from (\ref{solv}):
\begin{eqnarray}
\label{v2a}
\langle v^2\rangle (t)&=2\int_0^t d\tau (t-\tau) C_F(\tau) + 
2\int_0^t d\tau\int_0^t ds_1 \int_0^\tau ds_2 R(t-\tau) C_F(|s_1-s_2|)
\nonumber \\
  &+\int_0^t d\tau\int_0^t ds \int_0^\tau ds_1 \int_0^s ds_2 R(t-\tau)R(t-s) 
C_F(|s_1-s_2|).
\end{eqnarray}
Some of the above integrals have to be performed numerically. The result for
$T=1$ is presented in Fig.1. As expected, the system reaches the equilibrium
state. 

In the following, we will refer to the above result as "analytical". 
Alternatively, we can introduce a concrete stochastic process characterized by
covariance in the form (\ref{cor}) and calculate $\langle v^2\rangle$ from 
a Monte Carlo simulation. We apply the "kangaroo process" (KP). It is defined 
\cite{frisch} as a stepwise random function: $F(t)=F_i=\mbox{const}$ in the 
time interval $t_i \leq t < t_{i+1}$. The length of interval of constant $F$, 
$s$, is a function of the value of the process itself. The KP is a stationary
Markov process, determined by a stationary probability distribution 
$P_{KP}(F)$. It can be easily defined for arbitrary covariance. We get 
\cite{kan} the required form (\ref{cor}) by choosing 
$P_{KP}(F)=1/(2a)=\mbox{const}$ for $F\in (-a,a)$, zero elsewhere, where 
$a=\sqrt{3/\eta}$ with $\eta=\epsilon/T$. The time increment corresponding to 
a given $F$ follows from the formula $s=3a|F|^{-3}$. Thus $s$ assumes values 
between $\eta$ and infinity and its density distribution is of the form
\begin{equation}
\label{pods}
P(s)=(\eta^{1/3}/3)\; s^{-4/3}\; \theta (s-\eta),
\end{equation}
where $\theta(x)$ is the step function. Moments of $F$ can easily be obtained 
by averaging over the uniform distribution $P_{KP}(F)$. We get 
$\langle F\rangle=0$ and $\mbox{Var}\,(t)\equiv\langle F^2\rangle =1/\eta$.
Since $2P_{KP}(|F|)d|F|=P(s)ds$, we can average also over $P(s)$:
\begin{equation}
\label{f2}
\mbox{Var}\,(t)=3\eta^{-1/3}\int_\eta^\infty s^{-2/3}P(s)ds.
\end{equation}
Also the KP covariance \cite{frisch,kan} $C_{KP}$ can be expressed 
in terms of the distribution $P(s)$
\begin{equation}
\label{ckp}
C_{KP}(t) = 3\eta^{-1/3}
\int_\eta^\infty s^{-2/3} \exp(-t/s) P(s)ds.
\end{equation}
Inserting $F(t)$ into Eq.(\ref{solv}) allows us to determine the velocity 
time series of the Brownian particle. The variance at a given time $t$ is 
obtained simply by calculating $v(t)$, squaring it and averaging over many 
trajectories. Fig.1 presents the result: $\langle v^2\rangle$ does not 
stabilizes at the expected value $\langle v^2\rangle=T/m$ but instead it 
dwindles with time, obeying the approximate relation 
$\langle v^2\rangle(t) \sim t^{-0.67}$. 

The above outcome is surprising because the stationary process brings about 
an apparently non-stationary result. Moreover, according 
to (\ref{v2a}), $\langle v^2\rangle$ is completely determined by 
the covariance of $F$ and every simulation satisfying (\ref{cor}) 
should reproduce the analytical result. In order to
understand the origin of that inconsistency, let us reconsider in details 
how actually the stochastic force value enters the Langevin equation.
Evaluation of the Brownian particle velocity requires 
the value of $F$ at a {\it given} time $t$. For that purpose 
the distribution of $s$ is crucial because this value follows from 
the length of current interval in the stepwise evolution of KP. 
However, the requirement that we choose only those intervals which contain the
time $t$ imposes some bias, e.g. longer intervals are more probable. Therefore 
a distribution we actually use in the Langevin equation (the "effective" 
interval distribution), may not be identical with $P(s)$. Its cumulative 
distribution function, $\Phi(s,t)$, can be derived in the following way. First 
let us consider $s\leq t$ and assume that $t$ is found in $n+1$ interval, i.e. 
$S_n\equiv s_1+s_2+\dots +s_n<t$ and $S_{n+1}>t$. The probability that the sum 
of $n$ intervals yields a value between $x$ and $x+dx$ we denote by 
$P_n(x)dx$, providing that each component has the distribution $P(s)$. The 
distribution function $\Phi(s,t)$ is just equal to the conditional probability 
that an interval is larger than $t-x$, for any $x$ between $t-s$ and $t$, 
and any $n$ from 1 to $N$, where $N$ denotes the integer part of $t/\eta$: 
$\Phi(s,t)=\sum_{n=1}^N \int_{t-s}^t S_n(x) dx \int_{t-x}^s P(\xi) d\xi$.
Introducing $S(x)=\sum_{n=1}^N S_n(x)$ and inserting $P(x)$ from (\ref{pods}), 
we get the following equation:
\begin{equation}
\label{smt}
\Phi(s,t)=\eta^{1/3}\int_{t-s}^t S(x)[(t-x)^{-1/3}-s^{-1/3}]dx~~~~~\mbox{for}
~~~\eta\le s\leq t. 
\end{equation}
For $s>t$ the lower limit of integration extends to zero. Moreover, we have to
take into account also events for which $t$ is contained already in the first 
interval. The final formula reads:
\begin{equation}
\label{swt}
\Phi(s,t)=\eta^{1/3}\left\{\int_0^t S(x)[(t-x)^{-1/3}-s^{-1/3}]dx+
t^{-1/3}-s^{-1/3}\right\}~~~~~\mbox{for}~~~s>t. 
\end{equation}
The direct evaluation of $S(x)$ is very difficult. We can avoid it utilizing 
the normalization condition $\Phi(\infty,t)=1$. The function $S(x)$ must then 
satisfy the integral equation
\begin{equation}
\label{abel}
\int_0^t S(x)(t-x)^{-1/3}dx+t^{-1/3}=\eta^{-1/3},
\end{equation}
called Abel's equation. It possesses a weakly singular kernel, depending only
on the difference of its arguments. Therefore we can apply the Laplace 
transforms technique to solve it \cite{doe}. The solution is of the form
\begin{equation}
\label{sols}
S(x)=c_\Gamma\eta^{-1/3}\;x^{-2/3}-\delta(x),
\end{equation}
where a constant $c_\Gamma=1/[\Gamma(1/3)\Gamma(2/3)]\approx 0.2757\dots~$ 
contains the Gamma function. Inserting $S(x)$ to (\ref{smt}) and (\ref{swt})
and evaluating integrals gives us the expression for $\Phi(s,t)$. To obtain the
probability distribution ${\cal P}(s,t)$, we have to differentiate 
$\Phi(s,t)$ over $s$. The final result is simple:
\begin{eqnarray}
\label{pnods}
{\cal P}(s,t)=\cases {c_\Gamma [t^{1/3}-(t-s)^{1/3}]\;s^{-4/3}
& for $~~~\eta\le s\le t$\cr
(c_\Gamma t^{1/3}+\eta^{1/3}/3)\;s^{-4/3}& for $~~~~~~~~~\!s>t.$}
\end{eqnarray}

The effective interval distribution appears to be time-dependent and consisting
of two branches which do not join smoothly. We encounter a similar problem
asking about the mean time we must wait for a bus, providing we know 
the average time interval between subsequent bus arrivals ($\tau$). 
The answer is not $\tau/2$, as one could expect, but just $\tau$. 
This "waiting-time paradox" \cite{fel} can be elucidated by calculating
the effective, time-dependent probability distribution, analogous to 
(\ref{pnods}). In that case, however, the original distribution is an
exponential which results in the fast equilibrization and the effective 
distribution asymptotically becomes time-independent. For ${\cal P}(s,t)$ 
it never happens. Moreover, since the probability 
${\cal P}(s>t,t)=\int_t^\infty {\cal P}(s,t)ds=3c_\Gamma\approx 0.83$ 
does not diminish with time but remains constant, the entire distribution 
shifts with time towards long intervals. In fact, this outcome is not 
unexpected because all moments of ${\cal P}(s,t)$, as well as of $P(s)$, 
are divergent. On the other hand, long intervals correspond to small values 
of the process itself, which points out a reason of declining of the variance.
To derive expression for the effective variance ${\cal V}(t)$, we can use 
Eq.(\ref{f2}) substituting ${\cal P}(s,t)$ for $P(s)$. Evaluation of the 
integral gives us the final formula
\begin{equation}
\label{f2n}
{\cal V}(t)=c_\Gamma\eta^{-1/3}\;[3\ln3/2+\pi\sqrt{3}/6+
\ln(t/\eta)]\;t^{-2/3}~~~~~~~~~~(t\gg\eta).
\end{equation}
Hence the variance really decreases with time \cite{uwa1}. The KP covariance 
must also be modified. Replacing $P(s)$ in Eq.(\ref{ckp}) by 
${\cal P}(s,t_0)$, where $t_0$ is an initial time, and evaluating integrals 
we get the effective covariance
\begin{equation}
\label{ckp1}
{\cal C}_{KP}(t,t_0) = c_\Gamma\eta^{-1/3}
t_0^{1/3}t^{-1}\left[3-\exp(-t/2t_0)\Gamma(1/3)\;W_{-1/3,-1/2}\;(t/t_0)\right],
\end{equation}
where $W_{\alpha,\beta}\;(x)$ is the Whittaker function \cite{wit}. This result
is quite different from the original covariance (\ref{cor}) and explains us 
why the simulation does not agree with the analytical prediction (\ref{v2a}). 
${\cal C}_{KP}(t,t_0)\sim t^{-1}$ for large $t$ but it depends also on $t_0$. 

The above conclusions imply that problems involving algebraic correlations 
(e.g. critical phenomena, hydrodynamics), investigated in the framework of the
Langevin description, should be handled with caution. 
Conversely, an experimental evidence of declining variance in such
systems does not necessarily mean that Langevin formalism obeying standard
fluctuation-dissipation theorems does not apply. In any individual case one
should examine the distribution $P(s)$, the shape of which for large $s$ 
decides whether the system behaves in a stationary way. Is the stationary 
behaviour possible at all for correlations (\ref{cor})?  The analytical result 
would be valid for a steep $P(s)$. The fastest decaying distribution one can 
obtain with the KP for (\ref{cor}) declines asymptotically like $s^{-2}$ 
\cite{kan}. Since also for this distribution all moments diverge, we expect 
similar affects as for (\ref{pods}). 

Another possibility is to apply a deterministic process, instead of the 
Markovian stochastic one. For that purpose, let us consider a two-dimensional 
lattice of periodically distributed disks of radius $r$, with a particle 
bouncing elastically from them. Then the particle motion is free between 
subsequent collisions and its velocity ${\bf u}=(u_x,u_y)=\mbox{const}$. This
system, a periodic Lorentz gas, is equivalent to the Sinai billiard with 
periodic boundary conditions. We assume $2r<l$, where $l$ is the distance 
between disks centers. The system is strongly chaotic but the autocorrelation 
function of either component of particle velocity falls off slowly, as $1/t$ 
for large $t$ \cite{zach}. Therefore we can simulate solutions of (\ref{gle}) 
assuming the velocity of particle inside the independently evolved Sinai 
billiard as the stochastic force $F(t)$ \cite{sin}. A quantity of
interest is the distribution of free paths: it falls like $s^{-3}$\cite{bou},
steeper then for any KP. Its mean is convergent and the second moment weakly
divergent. For numerical simulations we assume $l=1$, $r=0.8$, $|{\bf u}|=1$ 
and $F=7.3\sqrt Tu_x$. Then $C_F=T/t$ for large $t$. We must stress, however, 
that the form of FAF at small $t$ also influences solutions of (\ref{gle}). 
Thus the simulations utilizing 
the Sinai billiard should be regarded as an approximation. The result of the
numerical calculation of the variance $\langle v^2\rangle(t)$ presents Fig.1. 
There are some discrepancies at small $t$, comparing to the analytical 
prediction, that can be attributed to differences in FAF. Asymptotically 
however, $\langle v^2\rangle$ stabilizes at the equilibrium value 
and both results coincide, in contrast to the KP case.

Finally, we wish to calculate the velocity autocorrelation function (VAF) 
$C_v(t)=\langle v(t_0)v(t_0+t)\rangle$, which is responsible for transport
properties of the system. It allows us, namely, to determine 
the diffusion coefficient ${\cal D}=\int_0^\infty C_v(t) dt$.
Typically, ${\cal D}$ is finite which corresponds to the normal diffusion. 
The analytical result for VAF in our case is \cite{glang}: 
$C_v(t)=T/m\;[1+\int_0^t R(\tau)d\tau]$. We present this function in Fig.2. 
It has the tail of the power-law shape with numerically estimated exponent 
equal to $-1.18$. On the other hand, we have calculated VAF from simulation 
utilizing the KP \cite{uwa}. The result for two values of $t_0$ is presented 
in Fig.2. We have normalized both functions to unity at $t=0$. Their shape is 
very different from the analytical result. The VAF initially falls but then it
stabilizes. It depends strongly on $t_0$, becoming more flat for larger $t_0$.
The stationary case applying the Sinai billiard also produces result different
from the analytical one -- $C_v(t)$ is always non-negative and does not 
approach zero for increasing time. As regards the transport properties 
of the system, the analytical $C_v(t)$ implies the normal diffusion. 
Determination of the precise shape of VAF at large $t$ for the KP and the 
simulation utilizing the deterministic random force, requires further studies. 
Anyway, it is obvious that the tails of VAF are very flat. Then the 
integration of VAF must produce a divergent result and $\cal D$ becomes 
infinite, leading to the diffusion process anomalously enhanced. This result 
agrees with that obtained in the framework of the continuous-time random walk 
approach \cite{kla} (L\'evi walks) predicting the enhanced diffusion 
for power-law distributions of free paths; for (\ref{pods}) the motion becomes 
ballistic: ${\cal D}$ diverges linearly with time.

\pagebreak

{\bf Figure captions}

FIG. 1. The velocity variance calculated from Eq.(\ref{v2a}) (solid line) 
and resulted from both simulations: with the KP (dashed line) and using the
deterministic random force (dots). The parameters: $T=1$, $m=1$ and
$\epsilon=0.01$. 

FIG. 2. The velocity autocorrelation function calculated using the KP with
$t_0=1.5$ (dot-dashed line) and $t_0=3$ (dashed line), normalized to unity at
$t=0$. The result of the simulation with the deterministic random force is 
marked by dots. The solid line shows the analytical result. The parameters 
are the same as in Fig.1.


\vskip2cm

\newpage

\begin{thebibliography}{99}
\begin{small}
\itemsep-2pt

\bibitem{maz}
R. M. Mazo, J. Chem. Phys. {\bf 54}, 3712 (1971).

\bibitem{cho}
T. S. Chow and J. J. Hermans, J. Chem. Phys. {\bf 56}, 3150 (1972).

\bibitem{cic}
B. Cichocki and B. U. Felderhof, J. Stat. Phys. {\bf 87}, 989 (1997).

\bibitem{frisch1}
A. Brissaud and U. Frisch, J. Quant. Spectrosc. Radiat. Transfer
{\bf 11}, 1767 (1971).

\bibitem{sron}
T. Srokowski and M. P{\l}oszajczak, Phys. Rev. Lett.
{\bf 75}, 209 (1995).

\bibitem{zach}
A. Zacherl, T. Geisel, J. Nierwetberg, and G. Radons,
Phys. Lett. {\bf 114A}, 317 (1986).

\bibitem{kubo}
R. Kubo, Rep. Prog. Phys. {\bf 29}, 255 (1966).

\bibitem{kubo1}
R. Kubo, M. Toda, and N. Hashitsume, {\it Statistical Physics II}
(Springer-Verlag, Berlin, 1985).

\bibitem{glang}
T. Srokowski and M. P{\l}oszajczak, Phys. Rev. E {\bf 57}, 3829 (1998).

\bibitem{frisch}
A. Brissaud and U. Frisch, J. Math. Phys. {\bf 15}, 524 (1974).

\bibitem{kan}
M. P{\l}oszajczak and T. Srokowski, Phys. Rev. E {\bf 55}, 5126 (1997).

\bibitem{doe}
G. Doetsch, {\it Anleitung zum praktischen Gebrauch der Laplace-Transformation}
(R. Oldenbourg, M\"unchen, 1956).

\bibitem{fel}
W. Feller, {\it An introduction to probability theory and its applications}
(John Wiley and Sons, New York, 1966), Vol.II.

\bibitem{uwa1}
The result (\ref{f2n}) does not mean, of course, that the KP is non-stationary
but only that it appears itself as such in the Langevin equation. 

\bibitem{wit}
E. T. Whittaker and G. N. Watson, {\it A course of modern analysis} (Cambridge
University Press, Cambridge, 1963).

\bibitem{sin}
M. P{\l}oszajczak and T. Srokowski, Ann. Phys. (N.Y.) {\bf 249}, 236 (1996);
T. Srokowski, Phys. Rev. E {\bf 59}, 2695 (1999).

\bibitem{bou}
J. P. Bouchaud and P. L. Le Doussal, J. Stat. Phys. {\bf 41}, 225 (1985).

\bibitem{uwa}
Since $C_v(t)$ can be expressed in a form similar to (\ref{v2a}) with 
${\cal C}_{KP}(t,t_0)$ instead of $C_F(t)$, our simulation resolves itself to
a Monte Carlo estimation of that multidimensional integral. 

\bibitem{kla}
G. Zumofen and J. Klafter, Physica D {\bf 69}, 436 (1993).

\end{small}

\end{thebibliography}
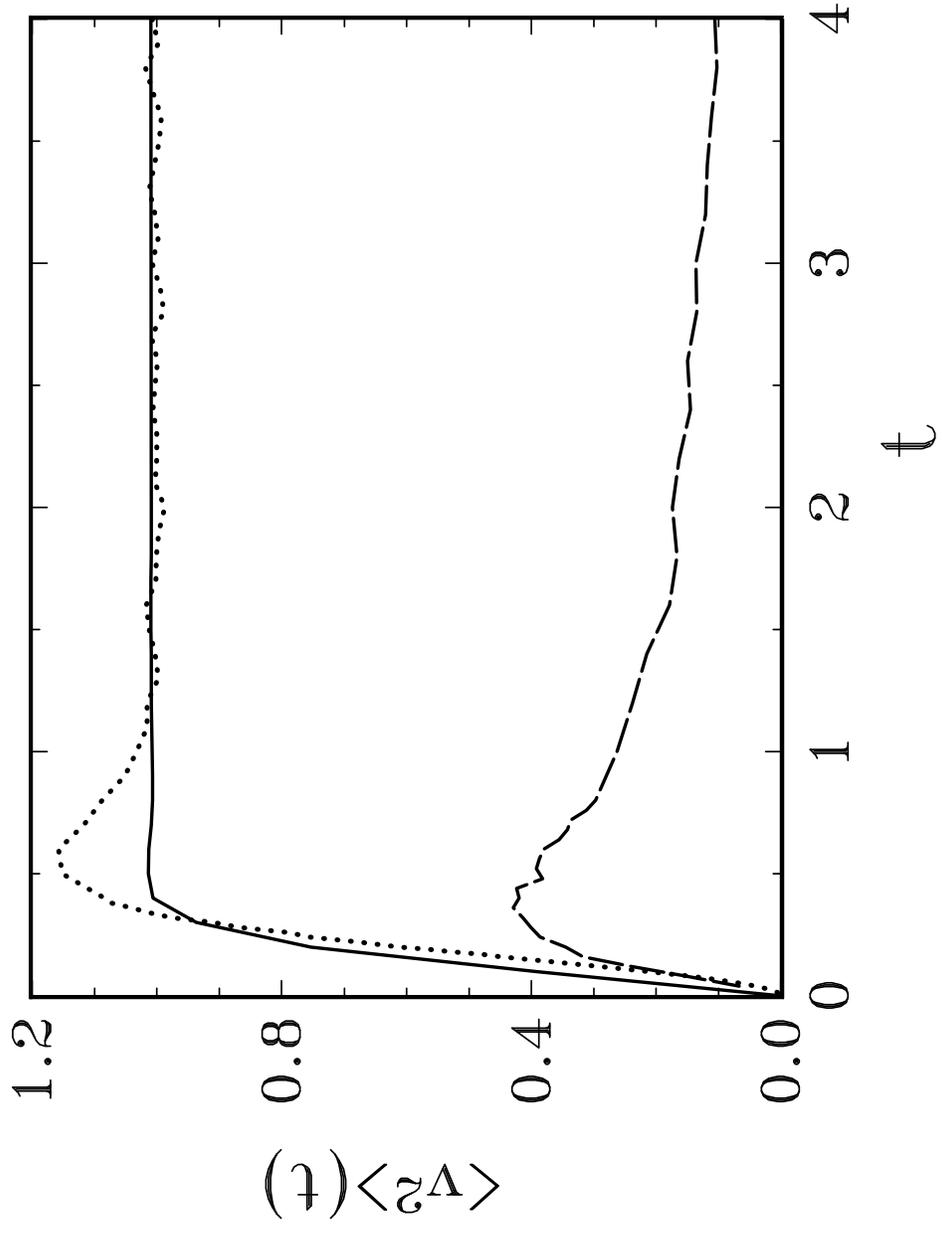
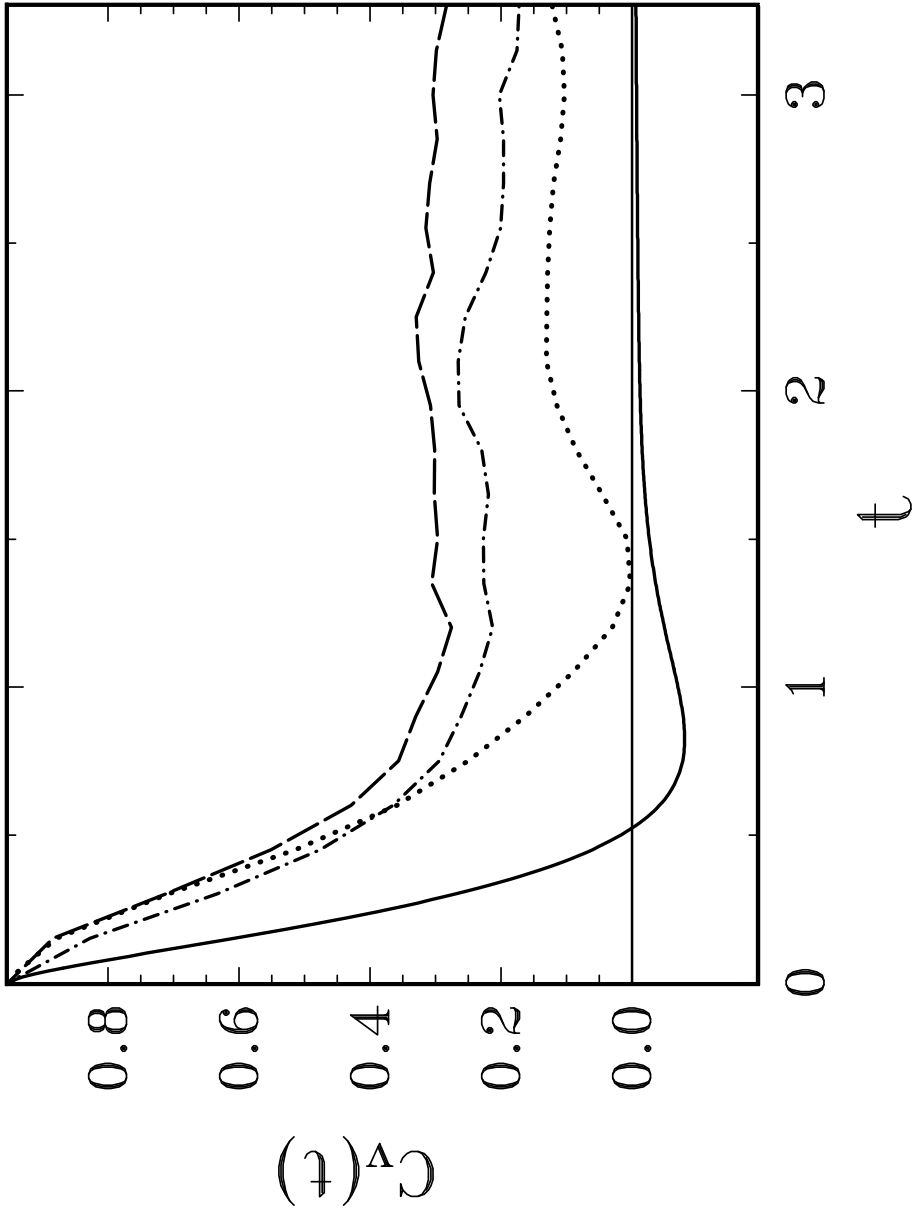
\end{document}